\documentclass{article} % For LaTeX2e
\usepackage{iclr2026_conference,times}

\usepackage{amsmath,amsfonts,bm}

\def\eqref#1{equation~\ref{#1}}
\def\1{\bm{1}}

\DeclareMathAlphabet{\mathsfit}{\encodingdefault}{\sfdefault}{m}{sl}
\SetMathAlphabet{\mathsfit}{bold}{\encodingdefault}{\sfdefault}{bx}{n}

\usepackage{url}
\usepackage{natbib}
\usepackage{booktabs} 
\usepackage{graphicx} 
\usepackage{caption}  
\usepackage{multirow} 
\usepackage{makecell}
\usepackage{subcaption}
\usepackage[dvipsnames]{xcolor}
\usepackage[colorlinks=true, citecolor=SkyBlue, linkcolor=red, urlcolor=black]{hyperref}

\title{Introducing Multimodal Paradigm for Learning Sleep Staging PSG via General-Purpose Model}

\author{Jianheng Zhou\textsuperscript{1},
    Chenyu Liu\textsuperscript{2}, 
    Jinan Zhou\textsuperscript{3}, 
    Yi Ding\textsuperscript{2}, 
    Yang Liu\textsuperscript{4}, 
    \textbf{Haoran Luo\textsuperscript{2}}, \\
    \textbf{Ziyu Jia\textsuperscript{5}}, \& 
    \textbf{Xinliang Zhou\textsuperscript{2}}\\
    \textsuperscript{1} School of Electrical and Electronic Engineering, Nanyang Technological University, Singapore \hspace{2pt} \\
    \textsuperscript{2} College of Computing and Data Science, Nanyang Technological University, Singapore\\
    \textsuperscript{3} Nutanix, CA, USA\hspace{2pt} 
    \textsuperscript{4}Center for Machine Vision and Signal Analysis, University of Oulu, \\Oulu, Finland\hspace{2pt}
    \textsuperscript{4}Institute of Automation, Chinese Academy of Sciences, Beijing, China \hspace{2pt} \\
    \texttt{Email: xinliang001@e.ntu.edu.sg}
}

\iclrfinalcopy % Uncomment for camera-ready version, but NOT for submission.
\begin{document}

\maketitle

\begin{abstract}
% Sleep staging is essential for diagnosing sleep disorders and assessing neurological health. However, existing automatic methods often extract features from complex and abstract raw polysomnograms (PSG) signal data and train domain-specific models, leading to a lack of data intuitiveness and strong reliance on a large scale of specific data. In this study, we propose an innovative paradigm for sleep staging based on a large multimodal general purpose model, aiming to mimic a doctor diagnostic process to achieve efficient and robust sleep staging. This method simplifies feature learning by converting the abstract one-dimensional time-series signals of PSG into intuitive and information-rich two-dimensional waveform images. Subsequently, a large multimodal general-purpose model is used to learn these PSG image representations. Extensive experiments on three datasets (ISRUC, MASS, and SHHS) demonstrate that a general-purpose model that has never seen relevant sleep data can acquire sleep staging capabilities by leveraging its inherent ability to learn fundamental and essential features. Our proposed method outperforms various existing state-of-the-art models.

Sleep staging is essential for diagnosing sleep disorders and assessing neurological health. Existing automatic methods typically extract features from complex polysomnography (PSG) signals and train domain-specific models, which often lack intuitiveness and require large, specialized datasets. To overcome these limitations, we introduce a new paradigm for sleep staging that leverages large multimodal general-purpose models to emulate clinical diagnostic practices. Specifically, we convert raw one-dimensional PSG time-series into intuitive two-dimensional waveform images and then fine-tune a multimodal large model to learn from these representations. Experiments on three public datasets (ISRUC, MASS, SHHS) demonstrate that our approach enables general-purpose models, without prior exposure to sleep data, to acquire robust staging capabilities. Moreover, explanation analysis reveals our model learned to mimic the visual diagnostic workflow of human experts for sleep staging by PSG images. The proposed method consistently outperforms state-of-the-art baselines in accuracy and robustness, highlighting its efficiency and practical value for medical applications. The code for the signal-to-image pipeline and the PSG image dataset will be released.
\end{abstract}

\section{Introduction}
Sleep staging is a critical process for diagnosing sleep disorders and monitoring neurological conditions. It is typically performed using polysomnography (PSG), which records multiple physiological signals such as electroencephalography (EEG), electrooculography (EOG), and electromyography (EMG). According to the American Academy of Sleep Medicine (AASM) standard \citep{berry2012rules}, experts divide the night's signals into 30-second segments, classifying them into five stages: W, N1, N2, N3, and REM. Manual scoring, however, is labor-intensive and subject to inter-scorer variability \citep{younes2016staging}, motivating the development of automated sleep staging methods.

%%%%%%%%%%%%%%%%%%%%%%%%%%%%%%%%%%%%%%%%%%%%
% Early automatic sleep staging research used machine learning \citep{pinero2004sleep,LAJNEF201594}, which depended on manually engineered features and struggled with high-dimensional data. To address this, deep learning method became popular. It automatically extracts features of a specific channel \citep{berthomier2007automatic,zhu2014analysis,mousavi2019deep,eldele2021attention} or a few combined channels \citep{8307462,jia2021salientsleepnet,10233098} for staging. However, these models exhibit limited robustness and generalization, as their performance degrades with the introduction of noise or under different data acquisition conditions. Inspired by these challenges and the success of large models, brain foundation models \citep{zhou2025brain} were introduced. These foundation models \citep{jiang2024large,wang2025cbramod,caro2024brainlm,dong2024brainjepa}, pre-trained on vast biomedical datasets, have largely overcome these issues and perform well across multiple tasks \citep{lai2025simple}.

%%%%%%%%%%%%%%%%%%%%%%%%%%%%%%%%%%%%%%%%%%%%
Early research applied traditional machine learning \citep{pinero2004sleep,LAJNEF201594}, relying on handcrafted features that struggled with the high dimensionality of PSG signals. Deep learning approaches have since achieved notable progress by automatically extracting features from individual \citep{berthomier2007automatic,zhu2014analysis,mousavi2019deep,eldele2021attention} or combined channels \citep{8307462,jia2021salientsleepnet,10233098}. Despite these advances, such models often suffer from limited robustness, with performance deteriorating under noisy conditions or across heterogeneous acquisition setups. Inspired by these challenges, recent efforts introduced brain foundation models (BFMs) \citep{zhou2025brain}, pre-trained on large-scale biomedical datasets, which improve generalization across tasks.

\begin{figure}[h!]
    \centering 
    
    % subgraph 1
    \begin{subfigure}{\linewidth}
        \centering
        \includegraphics[width=\linewidth]{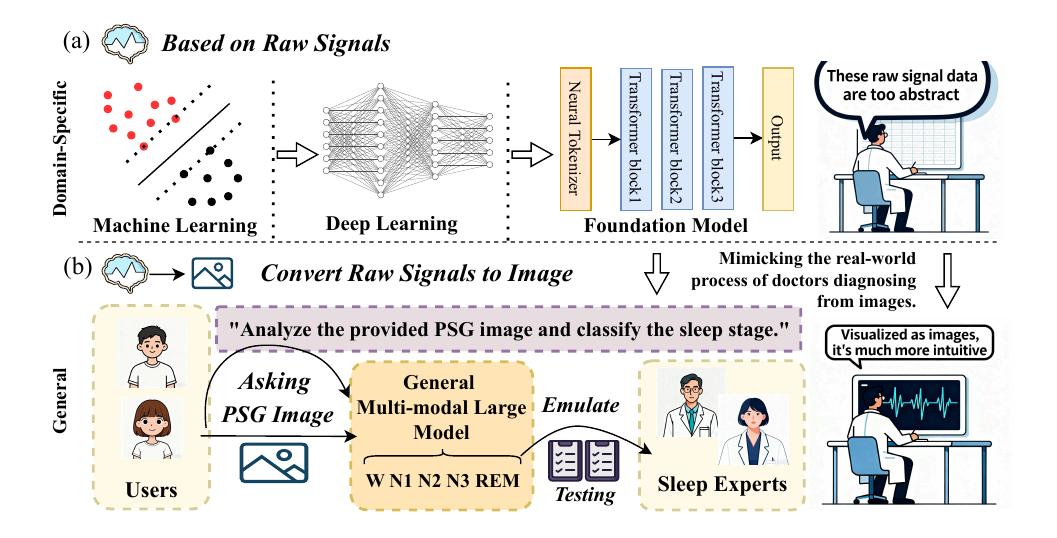}
    \end{subfigure}
    \par % 强制换行，确保下一个子图在下面
    \vspace{-1.5em} % 子图之间的垂直间距

    % subgraph 2
    \begin{subfigure}{\linewidth}
        \centering
        \includegraphics[width=\linewidth]{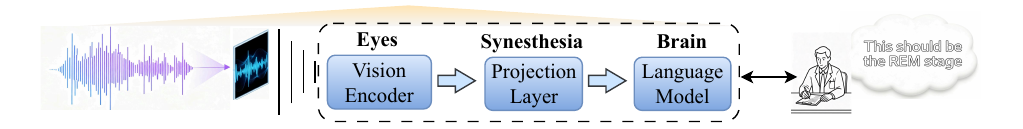}
    \end{subfigure}
    \caption{\textbf{(a)} illustrates the evolution of sleep staging research paradigms, from machine learning methods to more powerful deep learning, and finally to the brain foundation models that have emerged in recent years. \textbf{(b)} presents our proposed new paradigm, which breaks with the convention of using or training domain-specific models and relying on raw time-series signals, making the data more intuitive and the model more efficient and robust.} 
    \label{fig: intro}
\end{figure}

%%%%%%%%%%%%%%%%%%%%%%%%%%%%%%%%%%%%%%%%%%%%%
Nonetheless, two key limitations remain. \textbf{First, raw PSG signals are difficult to interpret and do not align with clinical practice}. Raw PSG signals are presented as a collection of one-dimensional time series from multiple channels. This abstract format is challenging to interpret, even for sleep experts, and becomes even more difficult when signals contain noise or artifacts. \textbf{Second, training domain-specific foundation models requires enormous amounts of in-domain data, making them costly and impractical for many applications.} The development of these models is highly demanding and has a strong dependence on vast amounts of specialized data. This issue is particularly acute in the medical domain, where data privacy concerns and the reliance on expert annotation make large-scale data acquisition both difficult and expensive. For instance, the pre-training data requirements of models like LaBraM \citep{jiang2024large} and CBraMod \citep{wang2025cbramod} are immense (approximately 2,500 hours from 20 datasets and a significantly larger 27,062 hours of EEG data, respectively). This scale of data underscores the prohibitively high cost associated with such specialized data collection.

Moreover, the prevailing research paradigm in sleep staging favors building domain-specific models from scratch. Without large-scale pre-training, the features learned by these models are often narrow and specialized, which may struggle to handle real-world variations in data distribution (e.g., different acquisition conditions). Despite demonstrating certain advantages in many studies, this paradigm overlooks the greater potential of large multimodal general-purpose models. During pre-training, these general-purpose models learn a universal understanding of modalities like images and text, and the broadly transferable representations they acquire can serve as a powerful foundation for applications in other specialized domains.
To overcome these challenges, we introduce a new paradigm for PSG-based sleep staging. \textbf{First, we design a signal-to-image conversion framework }that transforms abstract multi-channel time-series signals into intuitive waveform images, aligning the data representation with established clinical diagnostic workflows. This transformation bridges the gap between raw signal data and the visual format clinicians are trained to interpret. \textbf{Second, we leverage the large multimodal general-purpose model}, pre-trained on massive non-biosignal datasets, to exploit their broad capacity for learning transferable features. By fine-tuning these models with a relatively small set of PSG images, our approach achieves efficient and robust sleep staging.

Extensive experiments demonstrated our new paradigm to be a highly effective and versatile strategy for automated sleep staging. Our approach not only achieved better performance in key metrics such as accuracy, but also exhibited superior robustness when tested on data under noisy conditions. Furthermore, our findings show that by leveraging pre-trained knowledge, this paradigm can achieve state-of-the-art results with only a fraction of the data required by previous methods, proving its efficiency and practical value.

\section{Method}
\begin{figure}[t]
  \centering
  \includegraphics[width=\textwidth]{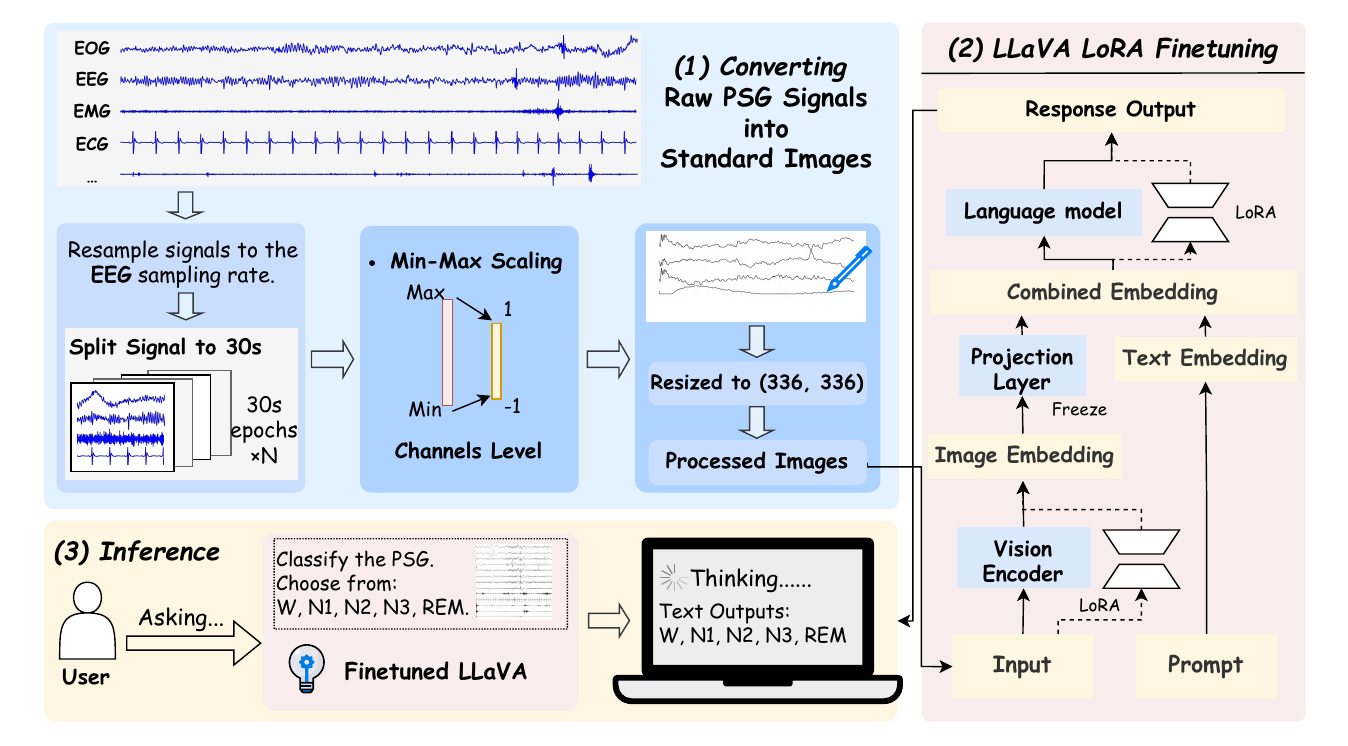}
  \caption{The Complete Workflow for the new PSG Image-based Sleep Staging Paradigm. \textbf{(1) Converting Raw PSG Signals into Standard Images.} First, we resample and segment the raw multi-channel PSG signals into 30-second epochs. Next, we scale each channel to a range of (-1, 1). Finally, the signals are plotted into images and resized to (336, 336). This standardized procedure ensures a consistent input format for the downstream model, making our pipeline reproducible across different datasets and research groups. \textbf{(2) LLaVA LoRA Finetuning.} The Vision Encoder and Language Model of LLaVA-Next were fine-tuned with LoRA, with the Projection Layer kept frozen. This approach allows us to equip the large, pre-trained LLaVA-Next model with the sleep staging capabilities without altering its vast foundational knowledge. \textbf{(3) Inference Validation on The Fine-tuned Model.} The model will output the sleep stage text corresponding to the input PSG image.}
  \label{fig: pipeline}
\end{figure}

The complete workflow of our new paradigm is shown in Fig. \ref{fig: pipeline}. First, we process the raw signal data through a standardized signal-to-image conversion pipeline to obtain image inputs. Next, we apply the large multimodal general purpose model, LLaVA-Next-7B model \citep{liu2024llavanext} and then use Low-Rank Adaptation (LoRA)  \citep{hu2022lora} to finetune it. Finally, we perform inference on the model with a variety of prompts to validate its classification capabilities.

\subsection{Signal Conversion Pipeline}
We define the \textbf{signal modality} as a multi-channel time-series matrix $S \in \mathbb{R}^{C \times T}$, where $C$ is the number of channels and $T$ is the number of time points. The \textbf{image modality} is a two-dimensional representation, $I \in \mathbb{R}^{H \times W}$, with dimensions $H \times W$. Our core methodology is an end-to-end signal-to-image conversion function $\mathcal{F}$, which includes all preprocessing, scaling, and rendering steps, and is expressed as: $I = \mathcal{F}(S)$.

\textbf{1) Unified Sampling Rate.}
First, signals from all available channels within each dataset are resampled to a unified EEG frequency, $f_{target}$ (e.g., 200 Hz in ISRUC). This ensures a consistent temporal baseline across all signals. Given an original discrete signal for a single channel, $s_{orig}[n]$, sampled at a frequency $f_{orig}$, the resampled signal $s_{resampled}[m]$ is obtained by a digital resampling process that can be expressed as:
\begin{equation}
   s_{resampled}[m] = \sum_{n=-\infty}^{\infty} s_{orig}[n] \cdot h\left(m \frac{f_{orig}}{f_{target}} - n\right) 
\end{equation}

Here, $h(t)$ is the impulse response of a low-pass interpolation filter. This operation is applied independently to each channel in the dataset.

\textbf{2) Signal Epoching.}
Following the resampling step, the continuous signals are partitioned into fixed-length epochs. Given a resampled signal matrix $S' \in \mathbb{R}^{C \times T_{\text{total}}}$ with $C$ channels and a total length of $T_{\text{total}}$ time points, the $i$-th epoch, $E_i$, is extracted as a sub-matrix of $S'$ with a fixed length of $N_{\text{epoch}}$ samples. This can be expressed as:
\begin{equation}
\label{eq:epoching}
E_i = S'[:, (i-1)N_{\text{epoch}} : i \cdot N_{\text{epoch}}]
\end{equation}

Here, $N_{\text{epoch}} = T_{\text{epoch}} \times f_{\text{target}}$ represents the number of samples in each epoch, where $T_{\text{epoch}}$ is the epoch duration in seconds (e.g., 30s) and $f_{\text{target}}$ is the unified sampling rate.

\textbf{3) Epoch-wise Min-Max Scaling.}
For each 30-second epoch, the data for each channel is independently processed by min-max scaling to the numerical range of [-1, 1]. Given a single epoch matrix $E_i$, the min-max scaling operation is applied to each channel $c$ independently. The scaled value for a sample $k$ in channel $c$, denoted as $E'_{i}[c,k]$, is calculated as:
\begin{equation}
\label{eq:minmax_scaling}
E'_{i}[c,k] = 2 \frac{E_i[c,k] - \min_{k'}\{E_i[c,k']\}}{\max_{k'}\{E_i[c,k']\} - \min_{k'}\{E_i[c,k']\}} - 1
\end{equation}

Here, $\min_{k'}\{E_i[c,k']\}$ and $\max_{k'}\{E_i[c,k']\}$ represent the minimum and maximum values of the $c$-th channel within the current epoch $E_i$, respectively.

\textbf{4) Image Rendering and Sizing.}
The complete data-to-image transformation is a \textbf{two-staged projection} from the normalized signal domain to the visual representation space. \textbf{First, high-Fidelity Rendering Phase.} The input epoch matrix $E'_{i} \in \mathbb{R}^{C \times T}$ is transformed into an initial high-resolution image $I'_{i, \text{initial}}$ by a \textbf{non-linear rendering functional} $\mathcal{R}_{\text{render}}$. This process spatially partitions each channel's data, ensuring non-overlapping placement of waveforms.
\begin{equation}
I'_{i, \text{initial}} = \mathcal{R}_{\text{render}}[E'_{i}] = \mathcal{F} \left( \bigoplus_{j=1}^{C} \mathcal{P}_j(E'_{i, j}) \right)
\end{equation}
Here, $\mathcal{P}_j$ is a channel-specific projection function, $\bigoplus$ is a non-linear superposition operator, and $\mathcal{F}$ is the final visual mapping functional. \textbf{Second, dimensionality Reduction Phase.} The intermediate image $I'_{i, \text{initial}}$ then undergoes a \textbf{down-sampling operator} $\mathcal{R}_{\text{size}}$ to produce a lower-dimensional tensor $I_i$ suitable for deep learning models. This involves a combination of spatial convolution and interpolation.
\begin{equation}
 I_i = \mathcal{R}_{\text{size}}[I'_{i, \text{initial}}] = \text{Conv}_{\downarrow} \ast \mathcal{L}_{\text{interp}}(I'_{i, \text{initial}})
\end{equation}
In this equation, $\text{Conv}_{\downarrow}$ is a down-sampling kernel and $\mathcal{L}_{\text{interp}}$ is an interpolation function. The final output is the model-ready tensor $I_i \in \mathbb{R}^{W_1 \times H_1}$.

We also explored the robustness advantages of converting signals into images, with the results shown in Section \ref{robust-eva} and the relevant proofs demonstrated in the Appendix \ref{Appn: sig-img-proof}.

% Finally, the processed epochs are rendered into images. This involves converting the one-dimensional, multi-channel signal data into a two-dimensional image format. Each channel waveform is plotted on a 2D canvas with a fixed pixel height (e.g., greater than 80 pixels) and a width equal to the number of samples in the epoch. The entire rendering and resizing process for a given epoch can be broken down into two sequential functions:
% \begin{gather}
% \label{eq:rendering}
% I'_{i, \text{initial}} = \mathcal{R}_{\text{render}}(E'_i) \\
% \label{eq:sizing}
% I_i = \mathcal{R}_{\text{size}}(I'_{i, \text{initial}})
% \end{gather}

% Here, $E'_i$ is the epoch matrix after min-max scaling. The first function, $\mathcal{R}_{\text{render}}$, renders the signal into an initial high-resolution image, $I'_{i, \text{initial}}$. The second function, $\mathcal{R}_{\text{size}}$, then resizes this initial image into the final $I_i$ tensor (e.g., $336 \times 336$ pixels) that is ready for the model input.

% It is important to note that the signal segments must not overlap. Each channel is rendered sequentially from top to bottom, with its waveform confined to a designated pixel area. While not strictly necessary, we recommend resizing the final image to $336 \times 336$, as this resolution aligns with the input requirements of most vision models and is computationally efficient.

\subsection{General-Purpose Model Finetuning}
In this work, we use \textbf{LoRA} to fine-tune the LLaVA-Next-Mistral-7B model for sleep staging. LoRA is applied to the individual linear layers of the vision encoder ($V$) and the language model ($L$), while the projection layer ($P$) remains frozen. This method efficiently adapts the model by representing the weight update for a layer input $x$ as a low-rank decomposition:
\begin{equation}
\label{eq:lora_weight_update}
h = Wx = W_0x + \frac{\alpha}{r}BAx
\end{equation}
The update term $\Delta W$ is decomposed into a low-rank matrix product of two smaller, trainable matrices, $B \in \mathbb{R}^{d \times r}$ and $A \in \mathbb{R}^{r \times k}$. The rank $r \ll \min(d,k)$, which significantly reduces the number of trainable parameters. A scaling factor $\alpha$ controls the magnitude of the update. This approach allows us to specialize the pre-trained LLaVA-Next model for our task of mapping visual features to a classification decision, with the full fine-tuned model represented by the forward pass:
\begin{equation}
\label{eq:lora_ft}
\hat{y} = L_{\text{LoRA}}(\text{prompt}, P(V_{\text{LoRA}}(I)))
\end{equation}

For implementation details, the training was performed with a learning rate scheduler that consists of a linear warmup phase followed by a cosine decay, to ensure stable convergence. These two phases are mathematically described as follows:
\begin{gather}
\label{eq:warmup}
\eta_t = \eta_{\text{max}} \cdot \frac{t}{T_{\text{warmup}}} \quad \text{for } t \in [0, T_{\text{warmup}}] \\
\label{eq:cosine_decay}
\eta_t = \eta_{\text{min}} + \frac{1}{2}(\eta_{\text{max}} - \eta_{\text{min}})\left(1 + \cos\left(\pi \frac{t - T_{\text{warmup}}}{T_{\text{max}} - T_{\text{warmup}}}\right)\right) \quad \text{for } t > T_{\text{warmup}}
\end{gather}

Here, $t$ is the current training step, $\eta_t$ is the learning rate at step $t$, $\eta_{\text{max}}$ is the peak learning rate (e.g., $2e-4$), $T_{\text{max}}$ is the total number of training steps, and $T_{\text{warmup}} = 0.1 \times T_{\text{max}}$.

Additionally, we investigated why large multimodal general-purpose model, when fine-tuned on unseen PSG images, outperform domain-specific models. The comparative results are presented in Section \ref{main-result}, and the theoretical proofs are detailed in the Appendix \ref{Appn: ft-proof}.

\subsection{Robustness and Explanation Analysis Methods}
\subsubsection{Robustness Analysis}
\label{low-quality}
Let $S$ be the set of all subjects in the dataset. We randomly selected a subset of subjects, $S_{aug} \subset S$, such that $|S_{aug}| \approx 0.70 \times |S|$. The remaining subjects, $S_{intact} = S \setminus S_{aug}$, were left unaltered. For each subject $s \in S_{aug}$ with a data recording $D_s$, we defined a random timestamp $t_{fail}$. The data $d_{s, c}(t)$ for a selected channel $c$ at time $t$ was then modified as follows:
\begin{equation}
    d'_{s, c}(t) = \begin{cases}
d_{s, c}(t) & \text{if } t < t_{fail} \\
\text{noise}(t) \text{ or } 0 & \text{if } t \geq t_{fail}
\end{cases}
\end{equation}

Here, $\text{noise}(t)$ represents white noise, and the choice between noise and zero values was made randomly. The random onset time $t_{fail}$ ensures that different subjects' data are affected to varying degrees.

To efficiently and pointedly validate the model robustness, we focused on a specific set of core EEG channels: $C_{core} = \{ \text{C3, C4, F3, F4, O1, O2} \}$. This selection is based on AASM guidelines and prior research confirming the importance of these channels for sleep staging.

\subsubsection{Explanation Analysis}
\label{FA}
Here, we used the \textbf{Feature Ablation} method, a perturbation-based technique, to understand our model sleep stage predictions from PSG images \citep{zhou2023interpretable}. This method systematically removes parts of the input to attribute the model decisions to specific image regions \citep{fisher2019all}.

We began by segmenting each PSG image into $576$ non-overlapping $14\times14$ pixel patches. For each patch $p_i$, a modified image $I_i'$ was created by masking the patch with a baseline value. The original model output is $\mathbf{y}(I) = f(I)$, and the output for the modified image is $\mathbf{y}(I_i') = f(M_i(I))$.

Next, we calculated an attribution score for each patch. The importance score $S(p_i)$ was determined by the Euclidean distance between the original and modified outputs:
\begin{equation}
S(p_i) = || \mathbf{y}(I) - \mathbf{y}(I_i') ||_2
\end{equation}

A higher score signifies greater importance. To ensure comparability, scores were normalized to $[0, 1]$ using min-max scaling:
\begin{equation}
 S_{norm}(p_i) = \frac{S(p_i) - \min(\{S(p_1), \dots, S(p_N)\})}{\max(\{S(p_1), \dots, S(p_N)\}) - \min(\{S(p_1), \dots, S(p_N)\})}
\end{equation}

The final normalized scores were used to generate an attribution heatmap, visually highlighting the most influential regions.

\section{Experiment}
\subsection{Datasets}
In this study, we tested our model on three publicly available PSG datasets:
\begin{itemize}
    \item ISRUC-Subgroup1 \citep{KHALIGHI2016180}, which follows the AASM staging standard.
    \item MASS-SS3 \citep{o2014montreal}, which follows the R\&K staging standard.
    \item SHHS-1 \citep{quan1997sleep,whitney1998reliability}, which follows the R\&K staging standard.
\end{itemize}
More detailed information about these datasets is provided in the Appendix \ref{dataset}.

\subsection{Experiment Setup}
For each dataset, we split the sample set into a train set and a test set with an 8:2 ratio. Subsequently, 10\% of the training set was randomly assigned as the validation set. To avoid bias from any specific data split, we performed this split five times using five different random seeds. The detailed hyperparameters for finetuning are shown in Appendix \ref{hyper}.

\subsection{Experiment Results}
\label{main-result}
For a comprehensive comparative analysis, we selected baseline models from three categories. The image-based classification models include: \textbf{ConvNext} \citep{Liu_2022_CVPR}, \textbf{MaxViT} \citep{tu2022maxvit}, \textbf{Resnet101} and \textbf{Resnet152} \citep{He_2016_CVPR}, \textbf{Swin-Transformer} \citep{Liu_2021_ICCV}, \textbf{ViT} \citep{dosovitskiy2021an}, and \textbf{Vision-Mamba} \citep{vim}. The signal-based classification models are: \textbf{MultiSleepNet} \citep{10146380}, \textbf{TinySleepNet} \citep{supratak2020tinysleepnet}, and \textbf{SleepWaveNet} \citep{10787148}. The EEG foundation models include: \textbf{LaBraM} and \textbf{CBraMod}. These baseline models cover a wide range of architectures and input types. Detailed descriptions of these comparative models and their training details are provided in the Appendix \ref{baseline} and \ref{setting}. The metrics used in the experiments are described in the Appendix \ref{Eva}.

\begin{table*}[htbp]
\centering
\caption{The results of different methods on sleep staging (ISRUC, MASS and SHHS).}
\label{tab:combined_results_full}
\setlength{\tabcolsep}{4pt}
\scalebox{0.74}{
\begin{tabular}{l l c c c c c c c c c c c c c c}
\toprule
& & \multicolumn{4}{c}{\textbf{ISRUC}} & \multicolumn{4}{c}{\textbf{MASS}} & \multicolumn{4}{c}{\textbf{SHHS}} \\
\cmidrule(lr){3-6} \cmidrule(lr){7-10} \cmidrule(lr){11-14}
\textbf{} & \textbf{Model} & \textbf{Acc} & \textbf{B-Acc} & \textbf{Kappa} & \textbf{W-F1} & \textbf{Acc} & \textbf{B-Acc} & \textbf{Kappa} & \textbf{W-F1} & \textbf{Acc} & \textbf{B-Acc} & \textbf{Kappa} & \textbf{W-F1} \\
\midrule
\multirow{7}{*}{\makecell{Image \\ based}} & ConvNext & 0.7558 & 0.7150 & 0.6815 & 0.7496 & 0.7761 & 0.6828 & 0.6616 & 0.7687 & 0.8074 & 0.7016 & 0.7300 & 0.8043 \\
 & MaxViT & 0.7875 & 0.7500 & 0.7231 & 0.7839 & 0.8281 & 0.7504 & 0.7436 & 0.8222 & 0.8474 & 0.7474 & 0.7863 & 0.8447 \\
 & Resnet101 & 0.7940 & 0.7528 & 0.7310 & 0.7907 & 0.8510 & 0.7959 & 0.7802 & 0.8482 & 0.8562 & 0.7574 & 0.7984 & 0.8535 \\
 & Resnet152 & 0.7958 & 0.7528 & 0.7330 & 0.7894 & 0.8475 & 0.7813 & 0.7718 & 0.8446 & 0.8590 & 0.7602 & 0.8023 & 0.8564 \\
 & Swin-ViT & 0.7551 & 0.7009 & 0.6779 & 0.7458 & 0.8007 & 0.7061 & 0.6985 & 0.7896 & 0.7734 & 0.6355 & 0.6769 & 0.7613 \\
 & ViM & 0.7759 & 0.7280 & 0.7064 & 0.7700 & 0.8118 & 0.7386 & 0.7197 & 0.8079 & 0.8382 & 0.7364 & 0.7738 & 0.8354 \\
 & ViT & 0.7022 & 0.6517 & 0.6108 & 0.6966 & 0.7710 & 0.6677 & 0.6533 & 0.7605 & 0.7348 & 0.6026 & 0.6228 & 0.7233 \\
\midrule
\multirow{3}{*}{\makecell{Signal \\ based}} & MultiSleepNet & 0.7169 & 0.6554 & 0.6278 & 0.7032 & 0.8079 & 0.7360 & 0.7178 & 0.8003 & 0.7815 & 0.7033 & 0.7014 & 0.7778 \\
 & TinySleepNet & 0.7896 & 0.7553 & 0.7422 & 0.7890 & 0.8321 & 0.7860 & 0.7741 & 0.8360 & 0.8584 & 0.7245 & 0.7996 & 0.8511 \\
 & SleepWaveNet & 0.7765 & 0.7121 & 0.6924 & 0.7128 & 0.8012 & 0.7293 & 0.7061 & 0.7161 & 0.8102 & 0.6971 & 0.7203 & 0.6927 \\
\midrule
\multirow{2}{*}{\makecell{Found- \\ ation}} & LaBraM & 0.7917 & 0.7503 & 0.7279 & 0.7886 & 0.8565 & 0.7965 & 0.7864 & 0.8544 & 0.8219 & 0.7216 & 0.7525 & 0.8207 \\
 & CBraMod & 0.8384 & 0.7891 & 0.7889 & 0.8290 & 0.8806 & 0.8065 & 0.8216 & 0.8720 & 0.8751 & 0.7613 & 0.8258 & 0.8714 \\
\midrule
\textbf{Ours} & \textbf{PSG-LLaVA} & \textbf{0.8596} & \textbf{0.8301} & \textbf{0.8172} & \textbf{0.8588} & \textbf{0.8879} & \textbf{0.8393} & \textbf{0.8337} & \textbf{0.8865} & \textbf{0.8782} & \textbf{0.7815} & \textbf{0.8291} & \textbf{0.8754} \\
\bottomrule
\end{tabular}
}
\end{table*}

In our model comparison, our proposed model significantly outperformed all others across every metric. Although the pre-trained EEG foundation models showed strong performance, they surprisingly held no clear advantage over general-purpose models, which had never been exposed to biosignal data before.

This finding has major implications for the medical field, given its common data scarcity challenges. The cost of collecting and pre-training domain-specific data for EEG models is far greater than that for general-purpose models. More importantly, we achieved superior performance with the general-purpose model using a much smaller dataset for fine-tuning. This suggests that these general models learn universally applicable features, allowing them to effectively adapt to and even outperform specialized models on new data, even with a significant distributional gap between their pre-training and fine-tuning datasets.

\subsection{Ablation Results}
\begin{figure}[htbp]
  \centering
  \includegraphics[width=1.0\textwidth]{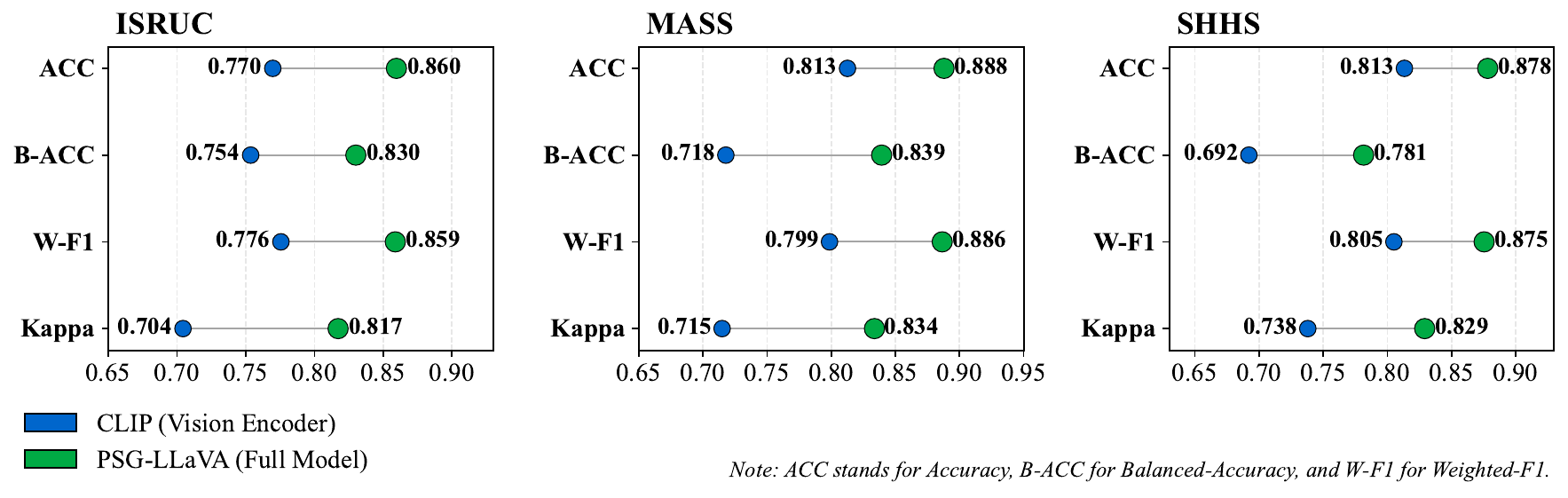}
  \caption{Comparison of Results between the Full LLaVA Model and its CLIP Vision Encoder.}
  \label{fig: CLIP-ablation}
\end{figure}

The rationale for choosing the complete LLaVA architecture is its powerful synergy between the vision encoder and language decoder. While CLIP \citep{radford2021learning} is a strong visual feature extractor, a standalone CLIP model, even with a modified classification head, underperformed compared to the complete LLaVA model (Shown in Fig. \ref{fig: CLIP-ablation}). LLaVA utilizes a pre-trained CLIP model as its robust vision encoder. However, a strong encoder alone is insufficient. The large language model (LLM) then acts as a powerful decoder, providing deep interpretation, contextual understanding, and reasoning. This allows the model to do more than just see the image; it understands the feature patterns and subtle differences, leading to enhanced robustness and higher accuracy for the sleep staging task. More supplementary ablation experiments can be found in Appendix \ref{abla-modal} and \ref{abla-ft}.

\subsection{Robustness Evaluation}
\label{robust-eva}
\begin{figure}[h!]   %htbp
  \centering
  \includegraphics[width=1.0\textwidth]{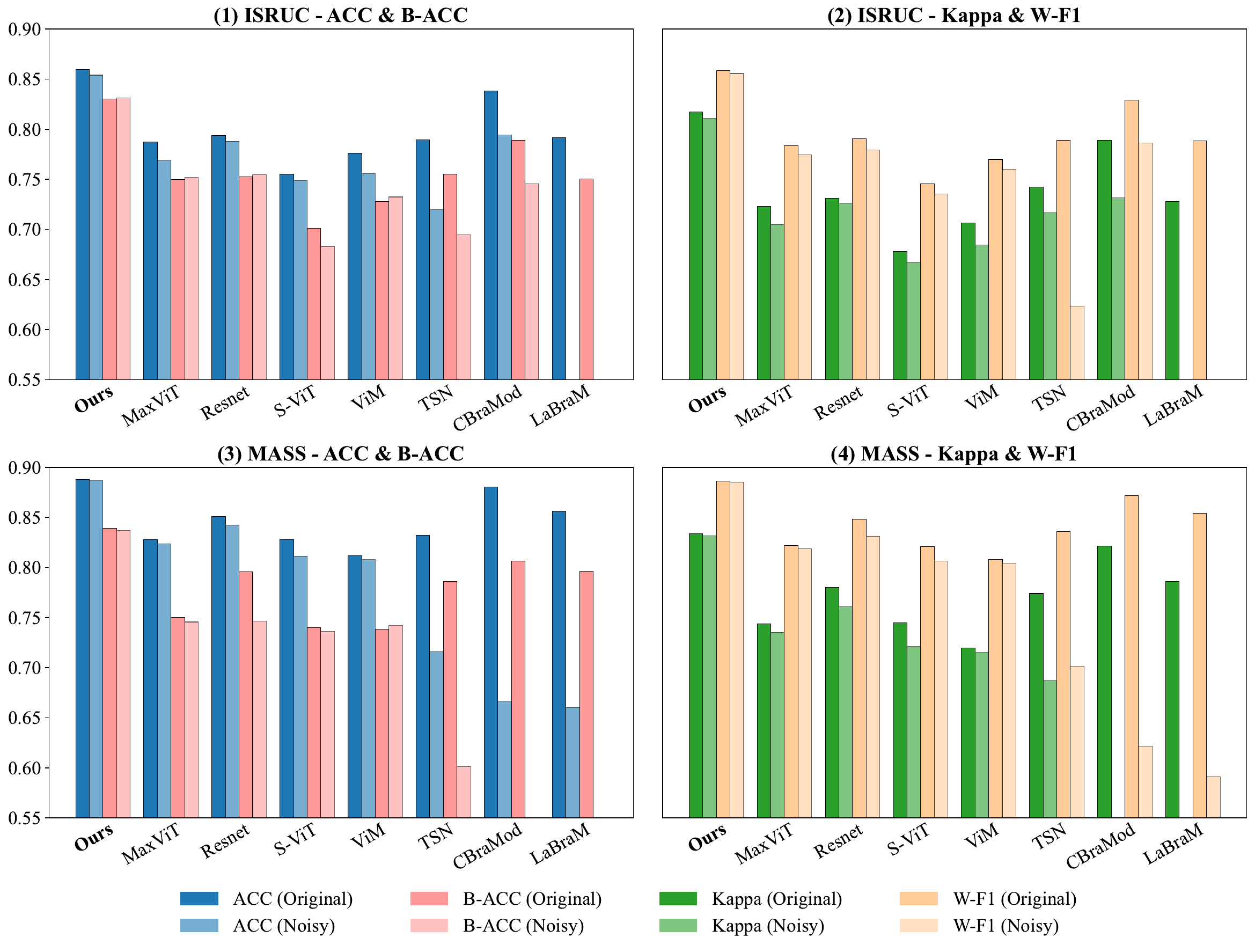}
  \caption{Model Performance Comparison Under Data Missing Challenges. \textbf{(1) and (2)} show the results for the ISRUC dataset. While \textbf{(3) and (4)} show the results for the MASS dataset.}
  \label{fig: robustness-1}
\end{figure}

To validate the model robustness, we processed the data using the method described in Section \ref{low-quality} to simulate real-world scenarios of poor channel quality or missing data. Models were trained and tested on this low-quality data with identical hyperparameter settings. While other models experienced a significant performance degradation, our model maintained excellent robustness across both datasets, with its performance staying at a high level, as shown in Fig. \ref{fig: robustness-1}.

\begin{figure}[h!]   %htbp
  \centering
  \includegraphics[width=1.0\textwidth]{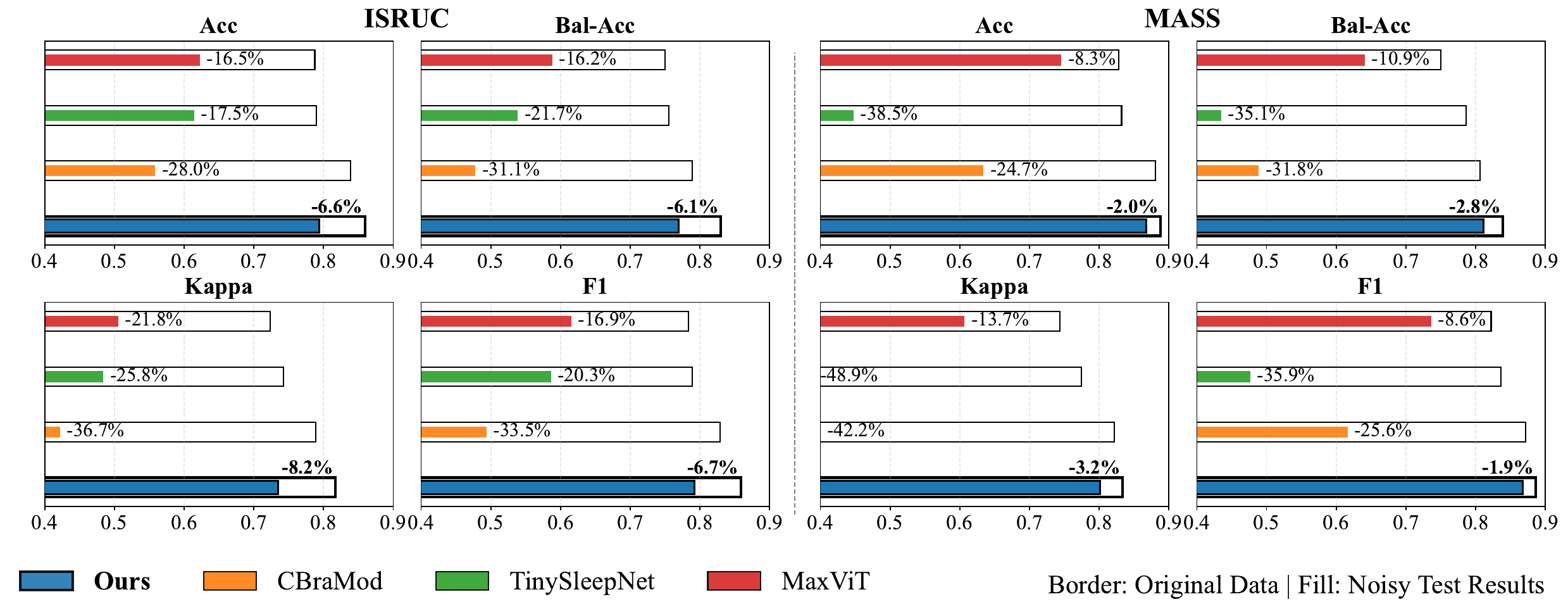}
  \caption{Robustness Test Performance on ISRUC and MASS. In the scenario where a model trained on normal data is tested on a low-quality dataset.}
  \label{fig: robustness-2}
\end{figure}

Moreover, we selected the best-performing model from each category for a second round of testing, simulating more realistic conditions. All four models were trained on normal data (lab environment) and tested on noisy data (real-world environment). We observed that our large multimodal general-purpose model maintained a high level of performance even under the influence of noise, as shown in Fig. \ref{fig: robustness-2}.

\subsection{Explanation Analysis}
To further interpret the internal mechanisms of our model for sleep staging, we conducted a feature ablation attribution analysis as Section \ref{FA} said. For better visual aesthetics, we applied some smoothing and retained only the top 30\% of patches based on their attribution importance to the final presentation. Through this approach, we can qualitatively determine which parts of the image are utilized by the model for sleep staging.

\begin{figure}[htbp]
  \centering
  \includegraphics[width=1.0\textwidth]{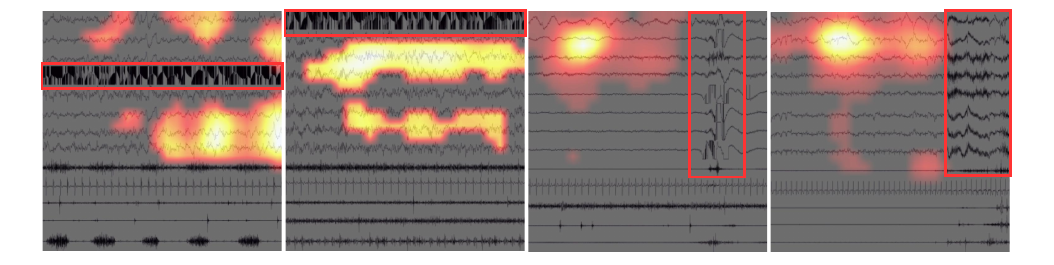}
  \caption{Explanation Analysis of Model Robustness on Anomalous Data. The signal within the red box represents a distorted signal affected by noise or artifacts.}
  \label{fig: heatmap_robust}
\end{figure}

To further explain the robustness of the large general-purpose model, attribution analysis was also performed on noisy samples. As shown in Fig. \ref{fig: heatmap_robust}, we can observe that the model does not treat low-quality channels or sudden bursts of noise as important features during inference. Instead, the model adaptively focuses on the uncontaminated parts of the signal. It is precisely for this reason that the model demonstrates significantly superior performance compared to other models in the robustness tests.

\begin{figure}[h!]   %htbp
  \centering
  \includegraphics[width=1.0\textwidth]{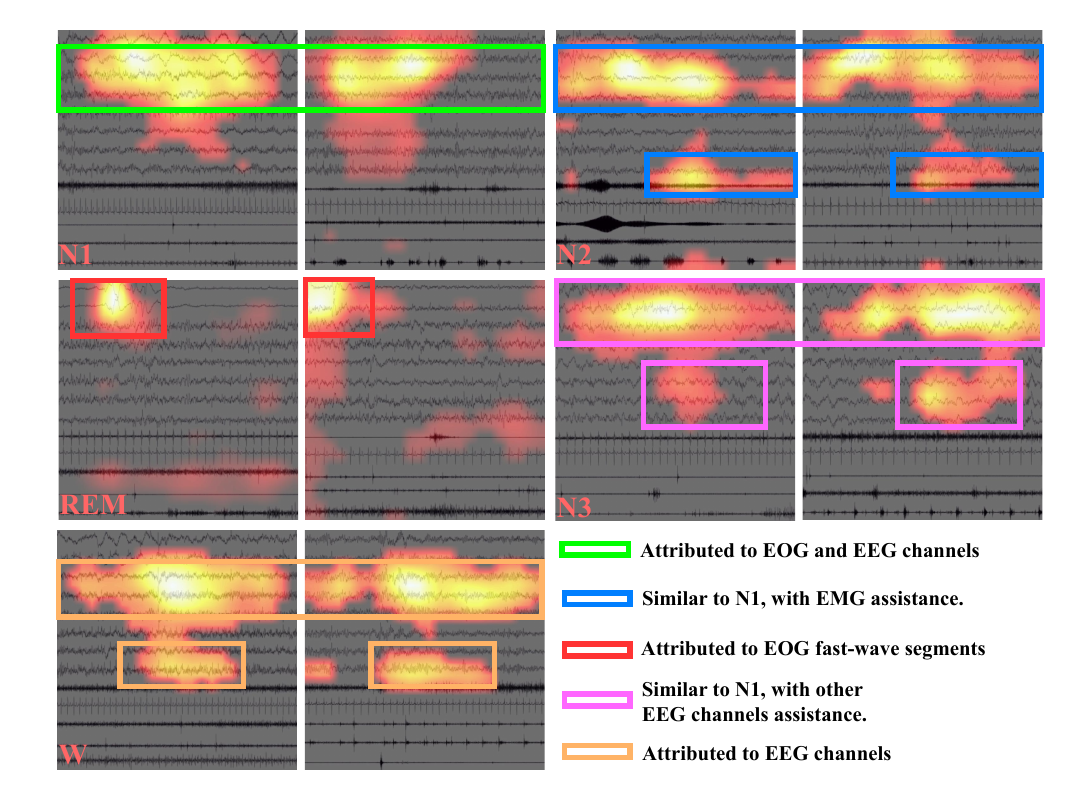}
  \caption{Explanation Analysis of Correctly classified Samples by Class. \textbf{N1:} The important regions primarily fall within the EOG and EEG channels, which the model uses collaboratively for inference. \textbf{N2:} Similar to N1, but the model also utilizes the assistance of the EMG signal. \textbf{REM:} The attribution analysis focuses on a specific signal segment: the rapid fluctuations in the EOG channel. \textbf{N3:} Similar to N1, and other EEG channels also provide auxiliary assistance. \textbf{W:} Primarily relies on all EEG channels.}
  \label{fig:combined_pdfs}
\end{figure}

The attribution results for a set of correctly classified samples are shown in Fig. \ref{fig:combined_pdfs}. In the attribution examples for the five sleep stages, we can observe that each category has its common attribution patterns. For instance, REM sleep attribution primarily focuses on the rapidly fluctuating signals in the EOG while other signals mainly serve an auxiliary role, and their significance is also evident in the figure to a certain extent. Explanation analysis for misclassified samples and for other datasets, can be found in the Appendix \ref{More FA}.

% \begin{figure}[h!]
%     \centering 
    
%     % 子图 1
%     \begin{subfigure}{\linewidth} % \linewidth 使子图占据当前文本宽度
%         \centering
%         \includegraphics[width=1.055\linewidth]{images/heatmap-N1R-compressed.pdf}
%     \end{subfigure}
%     \par % 强制换行，确保下一个子图在下面
%     \vspace{-0.3em} % 可选：子图之间的垂直间距

%     % 子图 2
%     \begin{subfigure}{\linewidth}
%         \centering
%         \includegraphics[width=1.053\linewidth]{images/heatmap-N2R-compressed.pdf}
%     \end{subfigure}
%     \par
%     \vspace{-0.3em}

%     % 子图 3
%     \begin{subfigure}{\linewidth}
%         \centering
%         \includegraphics[width=1.053\linewidth]{images/heatmap-N3R-compressed.pdf}
%     \end{subfigure}
%     \par
%     \vspace{-0.3em}

%     % 子图 4
%     \begin{subfigure}{\linewidth}
%         \centering
%         \includegraphics[width=1.048\linewidth]{images/heatmap-WR-compressed.pdf}
%     \end{subfigure}
%     \par
%     \vspace{-0.3em}

%     % 子图 5
%     \begin{subfigure}{\linewidth}
%         \centering
%         \includegraphics[width=1.066\linewidth]{images/heatmap-REMR-compressed.pdf}
%     \end{subfigure}

%     \caption{Explanation Analysis of Correctly classified Samples by Class. (The rows, from top to bottom, represent N1, N2, N3, W, and REM, respectively.)} % 整个大图的标题
%     \label{fig:combined_pdfs}
% \end{figure}

\section{Conclusion}
This study introduces a novel paradigm for automated sleep staging that mimics a doctor's diagnostic process. Our approach converts EEG signals into intuitive images and leverages a large multimodal general-purpose model (LLaVA) to achieve efficient and robust sleep staging.

Our results demonstrate that a general-purpose model, fine-tuned on a small amount of data, can surpass the performance of domain-specific models that undergo extensive pre-training. This finding offers a new solution for medical research fields facing data scarcity. Furthermore, our interpretability analysis shows that the model accurately simulates a doctor's diagnostic process by precisely focusing on key signal features. This characteristic is highly valuable for enhancing clinical trust and advancing AI-assisted medicine. Our work provides a new paradigm for EEG-based sleep staging and other signal processing tasks, with potential for broad application in various signal analysis fields.

% \subsection*{ETHICS STATEMENT}
% The data used in this study are derived from publicly available datasets for which usage rights have been granted. Data shown in this paper are represented in an aggregated, privacy-preserving format, such as heatmap overlays. We have carefully considered the potential ethical implications of this work and have concluded that there are no foreseeable risks or negative impacts associated with its use.

\subsection*{REPRODUCIBILITY}
We ensure the full reproducibility of all our experiments. To this end, our core code will be made publicly available once it has been organized.

\bibliography{iclr2026_conference}
\bibliographystyle{iclr2026_conference}

\clearpage
\appendix
% \section*{Use of LLMs}
% The authors confirm that the core research, methodology, and scientific findings presented in this paper were developed by the authors. Large Language Models (LLMs) were used exclusively for improving the clarity of the manuscript's language. No part of the scientific content, including research ideas, experimental design, or result analysis, was generated by an LLM. All figures were designed and laid out by the authors, with an LLM used solely to generate part of internal cartoon illustrations for aesthetic purposes.

\section{Related Work}
\label{R-Work}
Research on sleep stages has been conducted since a long time ago. Machine learning technologies like Support Vector Machines (SVM) \citep{hearst1998support} and Random Forests (RF) \citep{breiman2001random} are first applied in sleep staging area. To address the challenge in traditional machine learning that requires extensive prior knowledge \citep{liang2012automatic}, deep learning \citep{lecun2015deep} has been widely explored by researchers in the field of sleep staging. 
DeepSleepNet used Convolutional Neural Networks (CNN) and Bi-directional Long Short-Term Memory (BiLSTM) to extract the time-invariant features \citep{supratak2017deepsleepnet}. SleepEEGNet try to captures long short-term context dependencies by CNN with big scale kernel \citep{mousavi2019sleepeegnet}. XSleepNet model introduces a sequence-to-sequence approach that combines raw signals and time-frequency images \citep{phan2021xsleepnet}. To address model complexity, TinySleepNet was proposed as a less complex alternative to earlier models like DeepSleepNet \citep{supratak2020tinysleepnet}. Meanwhile, SleepUtime offers a completely feed-forward method that maps input sequences to class label sequences at a selected time resolution \citep{perslev2019u}. Some existing models provide interpretability, like Sleeptransformer \citep{phan2022sleeptransformer}, to better promote clinical trust. Graph neural networks (GNNs) \citep{4700287} have also shown advantages in the field of sleep stage classification by integrating the relationships between EEG channels. \citet{9530406} propose a multi-view spatial-temporal graph convolutional networks (MSTGCN) with domain generalization for sleep stage classification. There are more recent researches, like SleepWaveNet \citep{10787148}, which captures salient waves with inter-subject variability to improve subject adaptation. HSNN \citep{jia2022hybrid} and PicoSleepNet \citep{11139092}, which apply Spike Neural Network (SNN) \citep{ghosh2009spiking} for sleep staging using single-channel EEG signal.

Following the success of the Transformer architecture \citep{vaswani2017attention}, many transformer-based model emerge. ST-Transformer  \citep{song2021transformer} introduced multi-level transformers to encode spatial and temporal features across and within different channels. BIOT \citep{yang2023biot} is a prototype of biosignal foundation model. It innovatively tokenizes each signal channel independently and then flattens them into a unified "sentence" structure, thereby natively supporting signals with mismatched channels, variable lengths, and missing data. Recently, numerous studies on biosignal foundation models \citep{cui2024neuro,dong2024brainjepa,mckeen2025ecgfmopenelectrocardiogramfoundation,abbaspourazad2023large} pre-trained on large-scale biosignal data have also demonstrated excellent performance. LaBraM \citep{jiang2024large} is representative of this approach which is pre-trained on massive and diverse EEG datasets using masked signal modeling to learn widely transferable, general-purpose neural representations. Meanwhile, EEGPT \citep{wang2024eegpt} aims to be a "generalist model," employing an autoregressive pre-training task to enhance its adaptability across different devices and tasks. Furthermore, CBraMod \citep{wang2025cbramod} deeply optimizes the model architecture, introducing a "Criss-Cross" Transformer backbone that can model the inherent spatial dependencies and temporal dynamics of EEG signals separately, combining large-scale pre-training with a structure-aware network design.

\section{Datasets Introduction}
\label{dataset}
In our study, we introduced three famous datasets to support the research. They are ISRUC-Subgroup1, MASS-SS3, SHHS1. The detailed information shown below:

\textbf{ISRUC-Subgroup1 (ISRUC-S1):} The ISRUC-S1 dataset is a publicly available Polysomnography (PSG) dataset comprising complete overnight recordings from 100 adult subjects. The dataset is particularly valuable for its diversity, including data from healthy individuals, patients with various sleep disorders (with 62 cases of Obstructive Sleep Apnea Syndrome), and subjects under the effects of sleep medication. The recordings consist of six EEG channels, one ECG channel, two EOG channels, and three EMG channels, all sampled at 200 Hz. The sleep stages are expertly scored based on the American Academy of Sleep Medicine (AASM) guidelines, providing a robust resource for developing and evaluating automated sleep staging algorithms, especially for clinical applications.

\textbf{MASS-SS3:} The MASS-SS3 is a subset of the Montreal Archive of Sleep Studies (MASS), an open-access database aimed at providing a standardized benchmark for sleep analysis research. This subset contains full overnight laboratory-based PSG recordings from 62 subjects (29 males, 33 females) with an average age of 42.5 years. The dataset includes a comprehensive 21-electrode EEG montage, 2 EOG channels, 3 referential EMG channels, and 1 ECG channel, all sampled at 256 Hz. While the dataset is open-access, the biosignal files require submission of ethical approval to ensure subject privacy. This meticulous data collection and standardized format make MASS-SS3 an excellent resource for cross-validation and benchmarking of sleep staging models.

\textbf{SHHS1: }The SHHS1 dataset is the first phase of the large-scale Sleep Heart Health Study, a multi-center cohort study designed to investigate the cardiovascular consequences of sleep-disordered breathing. It includes PSG recordings from 6,441 men and women aged 40 and older. A key characteristic of this dataset is that the recordings were conducted in an unattended home setting, providing a unique "real-world" perspective on sleep behavior. The recordings include a variety of signals such as EEG (125 Hz), EOG (50 Hz), EMG (125 Hz), respiratory effort (10 Hz), and ECG (125 Hz or 250 Hz). With detailed annotations for sleep stages and respiratory events, SHHS1 is a valuable and widely used resource for studying the complex relationship between sleep and cardiovascular health. 

For our experiments, we selected the first 30 subjects in ISRUC, excluding subject 8 due to incomplete data, which resulted in the generation of 25,532 images. We selected the first 46 subjects but did not include subjects 40, 43, 45, and several subsequent ones due to issues with certain channels, generating a total of 40,976 images. For Sleep Heart Health Study (SHHS) dataset \citep{quan1997sleep,whitney1998reliability}, we selected data from 313 healthy subjects, generating a total of 314,676 images. As the public versions of these three datasets have already been filtered, we did not apply any additional processing to the data. Furthermore, we unified the staging standard across all datasets to AASM by merging the S3 and S4 stages defined in R\&K standard into N3.

\section{Finetuning Hyperparameters}
Table \ref{tab:hyperparams} shows the detailed hyperparams for finetuning. "Gradient steps" here refers to gradient accumulation steps. This parameter allows for the use of a larger effective batch size without requiring more GPU memory.
\label{hyper}
\begin{table}[h!]
\centering
\caption{Hyperparameters for LLaVA LoRA Fine-tuning}
\label{tab:hyperparams}
% 为了让换行单元格的左侧对齐更好看，我们将参数值的列设为 'l' (左对齐)
\begin{tabular}{l @{\hskip 0.8cm} l @{\hskip 1.5cm} l @{\hskip 0.8cm} l} 
\toprule
\textbf{Parameter} & \textbf{Value} & \textbf{Parameter} & \textbf{Value} \\
\midrule
LoRA Rank & 16 & Epochs & \makecell[l]{3.0/1.0(SHHS)} \\
Batch-size & 2 & Scheduler & cosine \\
Gradient steps & 8 & Warmup ratio & 0.1 \\
Learning rate & $1e-4$ & bf16 & true \\
\bottomrule
\end{tabular}
\end{table}

\section{Training Visualization}
Our model demonstrates rapid convergence, requiring only a small number of epochs for fine-tuning. The confusion matrix also illustrates key properties of its classification performance.
\begin{figure}[htbp]
  \centering
  \includegraphics[width=\textwidth]{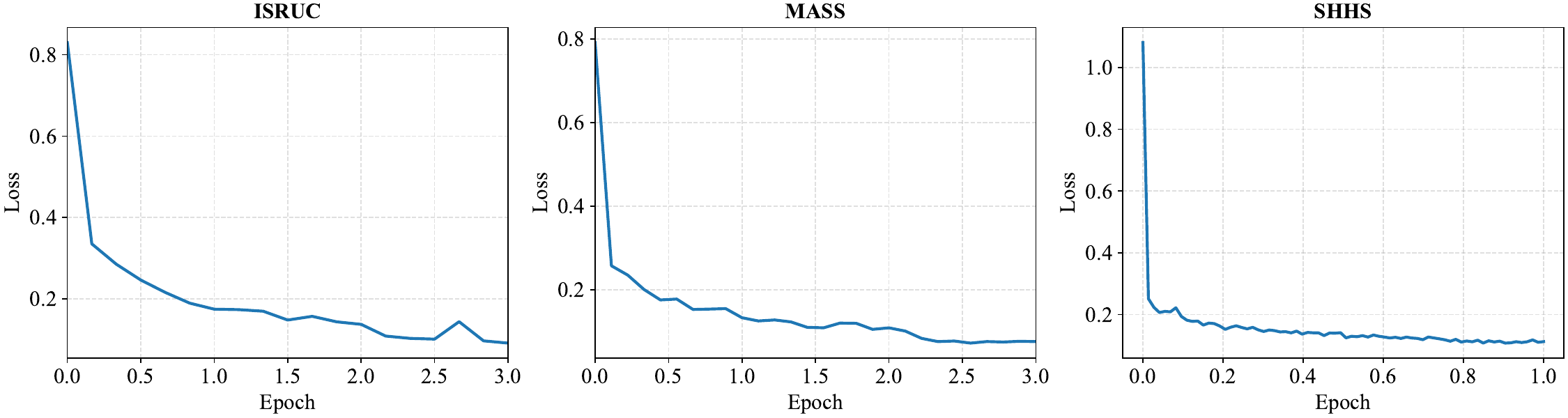}
  \caption{LLaVA Training Loss Curves on Three Datasets.}
  \label{fig: loss}
\end{figure}

\begin{figure}[htbp]
  \centering
  \includegraphics[width=\textwidth]{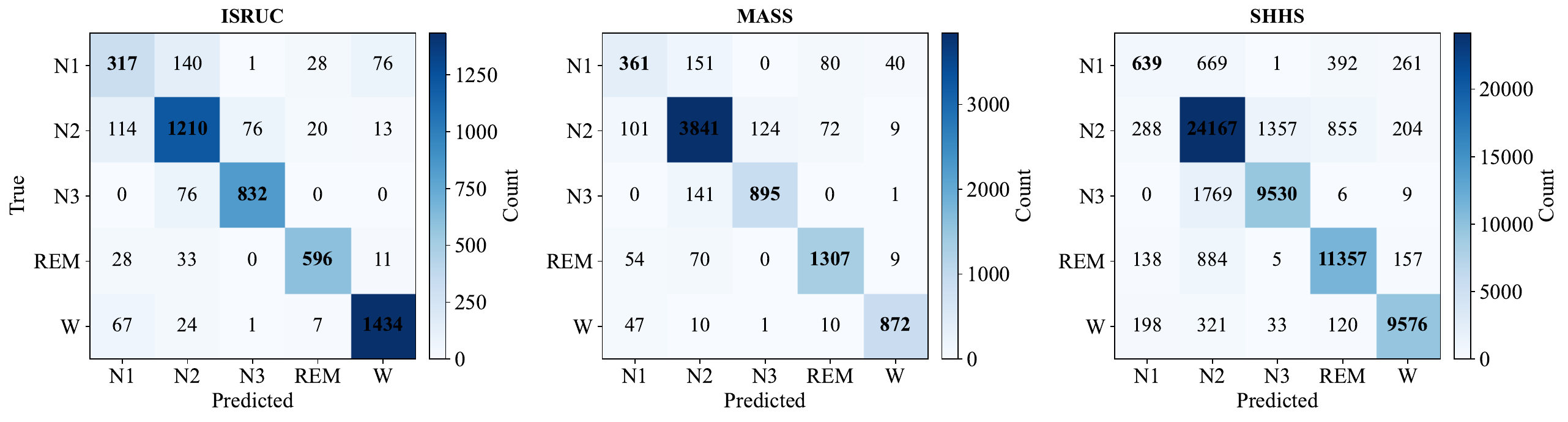}
  \caption{Average Inference Confusion Matrix on Three Datasets.}
  \label{fig: loss}
\end{figure}

\section{Baseline Introduction}
\label{baseline}
\textbf{ConvNeXt} reimagines the classic convolutional network architecture, modernizing its design by incorporating key insights from vision transformers. The model is a testament to the idea that pure convolutional networks can achieve competitive performance on par with transformer-based models on large-scale image datasets. By adopting a macro design, a simplified training regime, and specific architectural modifications from vision transformers, ConvNeXt demonstrates that convolutional networks are still highly scalable and effective for a wide range of computer vision tasks.

\textbf{ViT} (Vision Transformer) proposes an architecture that applies the transformer framework directly to image recognition tasks, challenging the long-standing dominance of convolutional neural networks. By dividing images into fixed-size patches and treating them as a sequence of tokens, the model leverages powerful self-attention mechanisms to learn global relationships and representations from a large-scale corpus of image data. This paradigm shift demonstrates that, with sufficient data, a pure transformer can outperform state-of-the-art CNNs, establishing a new foundation for various computer vision tasks.

\textbf{MaxViT} (Multi-axis ViT) proposes a hybrid architecture that seamlessly integrates the strengths of both vision transformers and convolutional neural networks. The model is built on a simple yet powerful design, using a combination of efficient multi-axis attention and convolutional layers. This innovative approach allows MaxViT to capture both local, fine-grained details and global, long-range dependencies in images, making it a highly effective and versatile backbone for a wide range of computer vision tasks.

\textbf{ResNet} (Residual Network) proposes a deep convolutional neural network architecture built upon the principle of residual learning. By introducing skip connections that allow gradients to bypass one or more layers, the model effectively addresses the vanishing gradient problem, enabling the training of exceptionally deep networks. This innovative framework allows ResNet to capture intricate, multi-scale features from large-scale image datasets, establishing it as a highly robust and versatile backbone for a wide range of computer vision tasks.

\textbf{Swin-ViT} (Swin Transformer) proposes a hierarchical transformer-based framework that addresses the computational scalability challenges of traditional vision transformers. By introducing a shifted windowing scheme, the model computes self-attention within local, non-overlapping windows, drastically reducing computational complexity while still enabling cross-window information exchange. This innovative approach allows Swin-ViT to capture both fine-grained, local features and global dependencies, making it a highly efficient and effective backbone for a wide range of computer vision tasks.

\textbf{ViM} (Vision Mamba) proposes an innovative architecture that combines the global context capabilities of transformers with the linear complexity of state-space models. By replacing the computationally intensive self-attention mechanism with a selective state space model, ViM captures long-range dependencies in a highly efficient manner. This breakthrough allows the model to process high-resolution images and long sequences with linear scalability, offering a compelling alternative to both transformers and convolutional networks while delivering competitive performance across a wide range of computer vision applications.

\textbf{MSNet} (MultiChannelSleepNet) proposes a Transformer encoder-based framework for automated sleep stage classification using multichannel PSG data. The model architecture is built on a Transformer encoder for both single-channel feature extraction and multichannel feature fusion. In the single-channel feature extraction module, a Transformer encoder independently extracts features from the time-frequency images of each channel. Based on an ensemble strategy, the feature maps extracted from each channel are fused in the multichannel feature fusion module. A second set of Transformer encoders further captures the joint features, while a residual connection in this module preserves the original information from each channel.

\textbf{TinySleepNet} proposes a lightweight convolutional neural network framework designed for accurate and efficient automated sleep staging. By incorporating a compact CNN architecture with a bidirectional Long Short-Term Memory (BiLSTM) layer, the model captures both local temporal patterns and long-range dependencies within a single EEG channel. This streamlined design makes TinySleepNet computationally efficient while maintaining competitive performance, establishing it as a highly practical solution for real-time and resource-constrained sleep analysis applications.

\textbf{SleepWaveNet} proposes a multimodal salient wave detection network, drawing inspiration from salient object detection tasks in computer vision. It employs a U-Transformer architecture to detect salient waves within sleep signals. Furthermore, a transfer learning-based individualized wave extraction framework adaptively extracts information specific to a target individual and identifies salient waves with inter-individual variability. The multimodal attention module can also adaptively enhance the importance of specific modal data for the sleep stage classification task.

\textbf{LaBraM} (Large Brain Model) proposes a scalable Transformer-based foundation model to address the lack of generic EEG representations from large-scale brain signal datasets. By pre-training on a diverse corpus of EEG recordings, the model captures rich temporal and spatial features that effectively transfer to various downstream BCI tasks. The architecture incorporates efficient self-attention mechanisms and task-specific adapters, enabling flexible fine-tuning while establishing a new paradigm for EEG signal analysis.

\textbf{CBraMod} (Criss-Cross Brain Foundation Model) proposes a Transformer-based EEG foundation model that addresses the complex and heterogeneous spatial and temporal dependencies inherent in EEG signals. It introduces a novel criss-cross attention architecture, comprising parallel spatial and temporal attention mechanisms, enabling separate yet simultaneous modeling of spatial and temporal relationships. This innovative framework allows the model to capture intricate, multi-dimensional EEG features, making it a robust and versatile backbone for a wide range of brain-computer interface (BCI) applications.

\textbf{CLIP} (Contrastive Language-Image Pre-training) proposes a powerful foundation model that addresses the complex and heterogeneous semantic relationships inherent in a massive amount of image-text pairs. It introduces a novel contrastive learning framework, comprising separate yet simultaneously trained image and text encoders, which project visual and linguistic data into a shared embedding space. This innovative approach allows the model to capture intricate, multi-modal features, making it a robust and versatile backbone for a wide range of zero-shot and open-vocabulary computer vision tasks.

\section{Evaluation Metrics}
\label{Eva}
To ensure a comprehensive and fair evaluation of the model performance, we employed the following four evaluation metrics:

\begin{align*}
% Accuracy (for multi-class)
\text{Accuracy} &= \frac{\sum_{c=1}^{C} TP_c}{N} \\
% Balanced Accuracy
\text{Balanced-Acc} &= \frac{1}{C} \sum_{c=1}^{C} \frac{TP_c}{TP_c + FN_c} \\
% Cohen Kappa
\kappa &= \frac{p_o - p_e}{1 - p_e} \\
% Weighted F1-Score
\text{Weighted-F1} &= \sum_{c=1}^{C} w_c \cdot F1_c
\end{align*}

% Preamble should include: \usepackage{amsmath} and \usepackage{amsfonts}
In the formulas above, the relevant terms are defined as follows. For any given class $c$:
\begin{itemize}
\item \textbf{True Positives ($TP_c$)}: The number of samples that are actually class $c$ and are correctly predicted as class $c$.
\item \textbf{True Negatives ($TN_c$)}: The number of samples that are not class $c$ and are correctly predicted as not being class $c$.
\item \textbf{False Positives ($FP_c$)}: The number of samples that are not class $c$ but are incorrectly predicted as class $c$.
\item \textbf{False Negatives ($FN_c$)}: The number of samples that are actually class $c$ but are incorrectly predicted as another class.
\end{itemize}

Furthermore, other parameters are defined as: $C$ is the total number of classes; $N$ is the total number of samples; $p_o$ is the observed accuracy; and $p_e$ is the expected chance agreement. For the Weighted F1-Score, the score for each class is the $F1_c$ score, calculated as $F1_c = 2 \cdot \frac{\text{Precision}_c \cdot \text{Recall}_c}{\text{Precision}_c + \text{Recall}_c}$. In this formula, $\text{Precision}_c$ is defined as $\frac{TP_c}{TP_c + FP_c}$ and $\text{Recall}_c$ is defined as $\frac{TP_c}{TP_c + FN_c}$. The final metric is a weighted average where the weight $w_c$ is the proportion of samples belonging to class $c$.

\section{Baseline Training Setup}
\label{setting}
Given that LLaVA-7B is a base version model, we default to using the base versions of other models for comparative analysis.

To ensure a fair comparison, all parameters for these general-purpose computer vision models were kept consistent, with the exception of ViT and Swin-ViT, which were trained for 50 epochs. For other models, only the epochs and batch sizes were slightly adjusted based on the loss trend and GPU memory constraints; all other parameters used their original settings. The detailed parameters shown as Table \ref{tab:training_config}.

\begin{table}[htbp]
  \centering
  \caption{Training Configuration for Different Models}
  \label{tab:training_config}
  \scriptsize
  \begin{tabular}{l >{\centering\arraybackslash}p{2.5cm} >{\centering\arraybackslash}p{4cm} ccccc}
    \toprule
    \textbf{Model} & \textbf{Strategy} & \textbf{Unfrozen Layers} & \textbf{B-Size} & \textbf{L-Rate} & \textbf{Weight-D} & \textbf{Epochs} \\
    \midrule
    \textbf{ConvNeXt} & Partial Finetuning & \makecell{Last stage/\\subsequent layers} & 64 & 1e-3 & 1e-4 & 30 \\
    \midrule
    \textbf{ViT} & Partial Finetuning & \makecell{Final three encoder blocks/\\classification head} & 64 & 1e-3 & 1e-4 & 50 \\
    \midrule
    \textbf{MaxViT} & Partial Finetuning & \makecell{Last stage/\\subsequent layers} & 64 & 1e-3 & 1e-4 & 30 \\
    \midrule
    \textbf{ResNet} & Partial Finetuning & \makecell{Layers 3\&4/\\classification head} & 64 & 1e-3 & 1e-4 & 30 \\
    \midrule
    \textbf{Swin-ViT} & Partial Finetuning & \makecell{Last stage/\\subsequent layers} & 64 & 1e-3 & 1e-4 & 50 \\
    \midrule
    \textbf{ViM} & Partial Finetuning & \makecell{Last stage/\\subsequent layers} & 64 & 1e-3 & 1e-4 & 30 \\
    \midrule
    \textbf{MSNet} & Full Training & / & 100 & 1e-3 & 1e-4 & 30 \\
    \midrule
    \textbf{TinySNet} & Full Training & / & 32 & 1e-4 & 1e-3 & 200 \\
    \midrule
    \textbf{SleepWNet} & Full Training & / & 8 & 1e-3 & 1e-4 & 10 \\
    \midrule
    \textbf{LaBraM} & Finetuning & / & 128 & 5e-4 & 5e-2 & 30 \\
    \midrule
    \textbf{CBraMod} & Finetuning & / & 32 & 1e-4 & 5e-2 & 30 \\
    \bottomrule
  \end{tabular}
\end{table}

\section{Modality Ablation Result}
\label{abla-modal}
\begin{figure}[htbp]
  \centering
  \includegraphics[width=\textwidth]{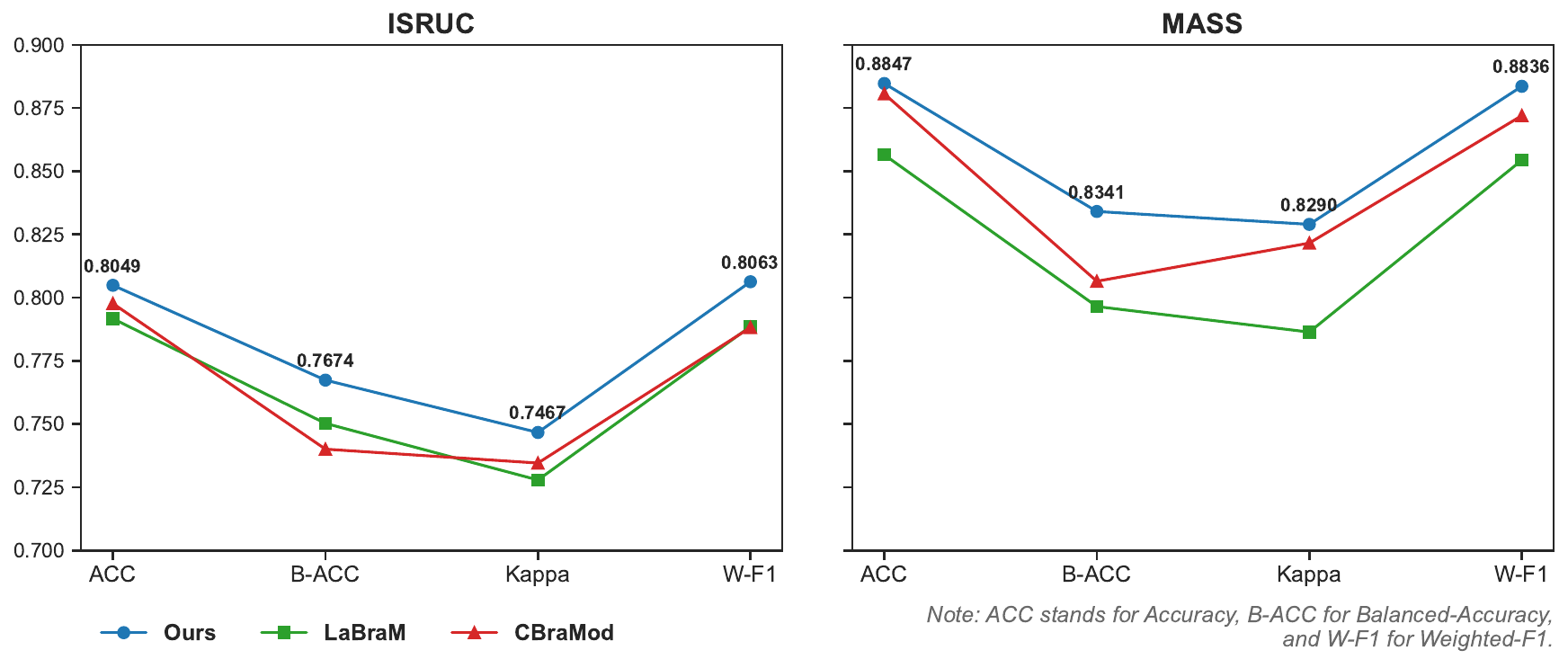}
  \caption{Comparison of Ablation Results for LLaVA Model Trained Exclusively on EEG Data with Foundation Model.}
  \label{fig: EEG-ablation}
\end{figure}

To ensure a fair comparison with the EEG-only foundation models, we conducted an ablation study to control for the advantage conferred by multimodality. In this experiment, we removed all non-EEG channels and evaluated our model performance on single-modality (EEG) data alone. The results are presented in Fig. \ref{fig: EEG-ablation}. We can see that despite a performance drop in ISRUC, our model still outperforms two EEG-based models on both datasets when using only single-modality EEG data.

\section{Ablation Study on Fine-tuning}
\label{abla-ft}
LLaVA is already a powerful multimodal large language model for images and text, and its capabilities in image understanding and text generation are well-established. Although it has never been exposed to medical images, particularly PSG waveform plots, we used the original, untuned LLaVA model for inference and performance testing to more rigorously demonstrate the effectiveness of our research workflow.

\begin{figure}[htbp]
  \centering
  \includegraphics[width=\textwidth]{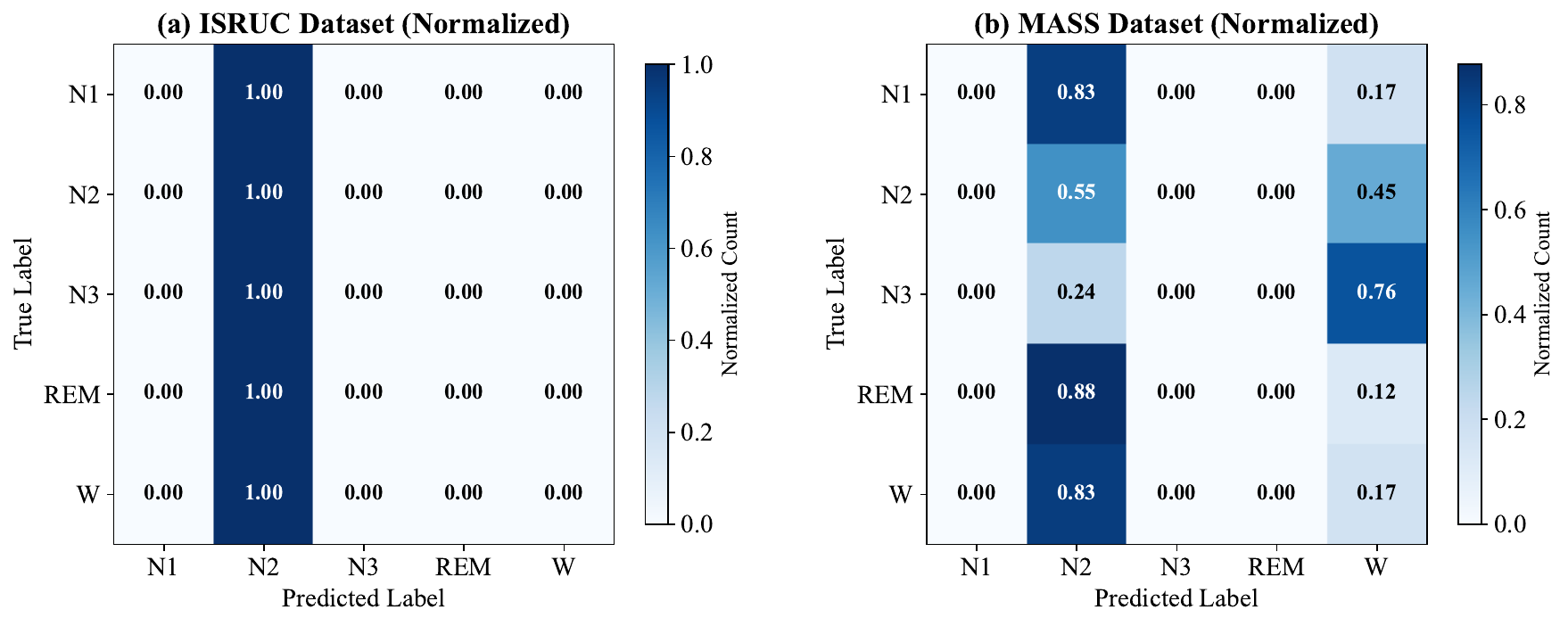}
  \caption{Confusion Matrix for LLaVA Model without Finetuning in ISRUC and MASS Dataset.}
  \label{fig: ft-ablation}
\end{figure}

The results are shown in the Fig. \ref{fig: ft-ablation}. When using the original, untuned LLaVA model for sleep stage inference, all samples are classified into one or two classes, especially stage N2, which is the most numerous. This indicates that the original model is not capable of identifying sleep stages based on PSG images, which is quite intuitive.

% \section{More Robustness Analysis}
% \label{robust-analysis}
% To conduct a robustness test under more challenging conditions, we used models trained on the original data and evaluated them on a poor-quality test dataset. We selected the best-performing models from the robustness tests in Section \ref{robust-eva} for each category: MaxViT, TinySleepNet, and CBraMod. The test results are shown in the Fig. \ref{fig: ISRUC_robust} and \ref{fig: MASS_robust}, where it can be seen that, besides our model, the performance of the other three models significantly dropped. This further confirms the high robustness of our proposed model.

% \begin{figure}[htbp]
%   \centering
%   \includegraphics[width=\textwidth]{images/ISRUC_robust_comparison.pdf}
%   \caption{\textbf{Robustness Test Performance on ISRUC.} In the scenario where a model trained on normal data is tested on a low-quality dataset.}
%   \label{fig: ISRUC_robust}
% \end{figure}

% \begin{figure}[htbp]
%   \centering
%   \includegraphics[width=\textwidth]{images/MASS_robust_comparison.pdf}
%   \caption{\textbf{Robustness Test Performance on MASS.} In the scenario where a model trained on normal data is tested on a low-quality dataset.}
%   \label{fig: MASS_robust}
% \end{figure}

\section{Explanation Analysis}
\label{More FA}
\begin{figure}[h!]  %htbp
  \centering
  \includegraphics[width=0.9\textwidth]{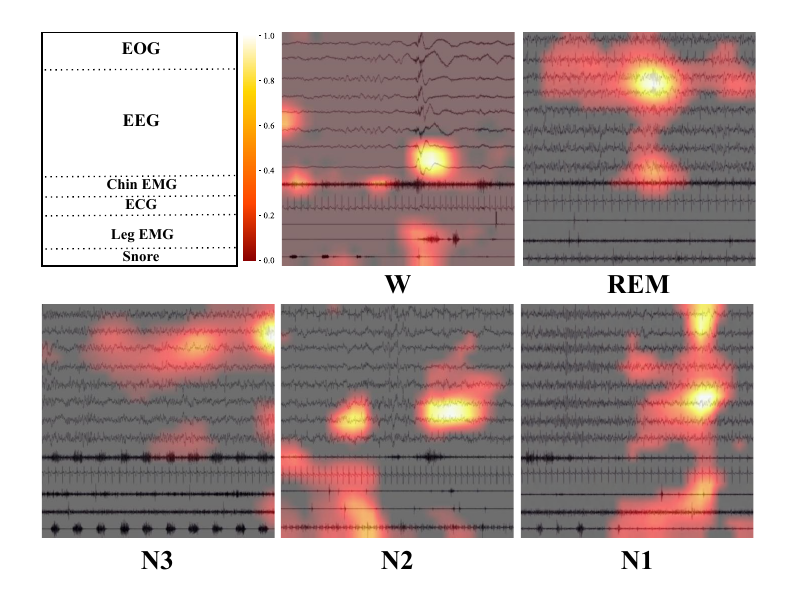}
  \vspace{-1.5em}
  \caption{Explanation Analysis of Misclassified Samples}
  \label{fig:wrong-heatmap}
\end{figure}

Evidently, the attribution patterns of misclassified samples (shown in Fig. \ref{fig:wrong-heatmap}) deviate significantly from the general pattern shown in Fig. \ref{fig:combined_pdfs}. Moreover, the regions highlighted as important do not fall within the signal areas that experts would typically use to determine sleep stages. This accounts for the model failure, yet it also indirectly reflects the consistency of its inference.

\begin{figure}[h!]
    \centering 
    
    % subgraph 1
    \begin{subfigure}{\linewidth}
        \centering
        \includegraphics[width=\linewidth]{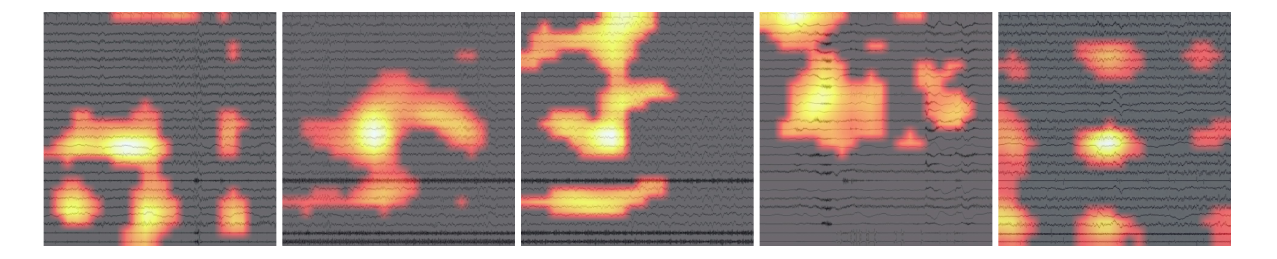}
    \end{subfigure}
    \par % 强制换行，确保下一个子图在下面
    \vspace{-1em} % 子图之间的垂直间距

    % subgraph 2
    \begin{subfigure}{\linewidth}
        \centering
        \includegraphics[width=\linewidth]{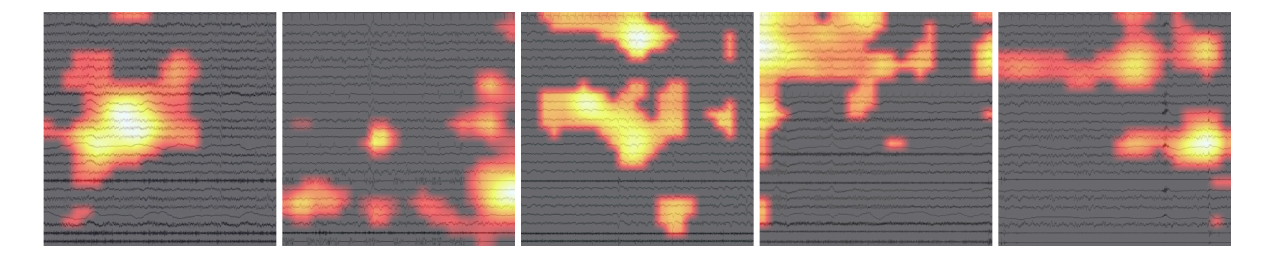}
    \end{subfigure}
    \par
    \vspace{-1em}

    % subgraph 3
    \begin{subfigure}{\linewidth}
        \centering
        \includegraphics[width=\linewidth]{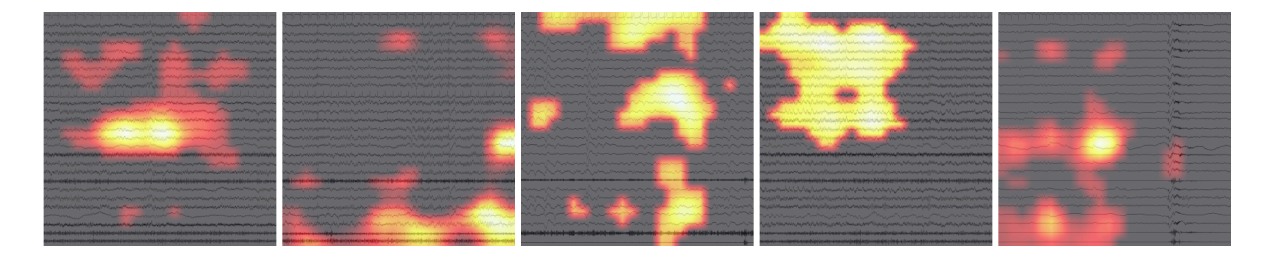}
    \end{subfigure}

    \caption{Explanation Analysis of Correctly classified Samples in MASS Dataset by Class. (The columns, from left to right, represent N1, N2, N3, W, and REM, respectively. Signal order from top to bottom: ECG, EEG-F8, EEG-O2, EEG-P3, EEG-C4, EEG-T5, EEG-CZ, EEG-P4, EEG-C3, EG-A2, EEG-F7, EEG-Fp1, EEG-T6, EEG-F4, EEG-FZ, EOG-R, EEG-T4, EEG-F3, EEG-OZ, EMG-Chin2, EEG-Fp2, EEG-PZ, EEG-O1, EOG-L, EEG-T3, EMG-Chin1, EMG-Chin3)} 
    \label{fig:mass}
\end{figure}

An attribution analysis on MASS (Fig. \ref{fig:mass}) shows that with a high number of channels, the model tends to rely on more signals to make its decisions. As a result, even though the overall attribution patterns are similar, many different auxiliary areas of importance are also highlighted. Notably, the model still focuses most on the critical information. Taking REM sleep as an example, despite the EOG channels only accounting for two of the total, the attribution results show that the important regions still fall on these two channels. This demonstrates that the model does not overlook crucial information hidden in dense channels.

\end{document}